# Analysing of 3D MIMO Communication Beamforming in Linear and Planar Arrays

Amirsadegh Roshanzamir
Department of Electrical Engineering
Sharif University of Technology
Tehran, Islamic Republic of Iran
roshanzamir_as@alum.sharif.edu

*Abstract*— Massive multiple-input multiple-output (MIMO) systems are expected to play a crucial role in the 5G wireless communication systems. These advanced systems, which are being deployed since 2021, offer significant advantages over conventional communication's generations. Unlike previous versions of communication, MIMO systems can transmit various probing signals through their antennas, which may or may not be correlated with each other. This waveform diversity provided by MIMO communication enables enhanced capabilities and improved performance.

Numerous research papers have proposed different approaches for beamforming in MIMO communication. We anticipate that our research will provide valuable insights into the performance of different beamforming techniques for MIMO communication systems with planar arrays. We will investigate the 3D beam patterns generated by these constellations using the covariance-based MIMO communication waveform method. MATLAB simulations will be utilized to analyze and evaluate the performance of these methods.

## I. INTRODUCTION

In the realm of wireless communication systems, the ever-increasing demand for higher data rates, improved reliability, and efficient spectrum utilization has necessitated the development of advanced transmission techniques. Multiple-Input Multiple-Output (MIMO) technology has emerged as a promising solution to address these challenges by exploiting the spatial dimension of wireless channels.

MIMO communication systems employ multiple antennas at both the transmitter and receiver ends, enabling the simultaneous transmission and reception of multiple data streams. By leveraging the spatial diversity provided by these antennas, MIMO systems offer significant advantages over traditional single-antenna systems, including increased capacity, improved link reliability, and enhanced spectral efficiency [1, 4, and 5].

A crucial aspect of MIMO communication lies in the design of optimal beampatterns. A beampattern represents the directional sensitivity of a MIMO antenna array, determining how the transmitted or received signals are spatially distributed. Properly designed beampatterns can significantly enhance system performance by focusing the transmitted energy towards desired directions while mitigating interference from unwanted directions [7, 8, 9, and 10].

The design of MIMO beampatterns is a multidimensional optimization problem involving several key factors. These factors include antenna geometry, array configuration, signal processing algorithms, and channel characteristics. Achieving an optimal beampattern requires careful consideration of these factors to maximize the desired signal power while minimizing interference and maintaining compatibility with existing communication standards [11, 12].

This paper aims to explore the various aspects involved in MIMO communication beampattern design. It will delve into the fundamental principles of MIMO systems, highlighting the benefits of exploiting spatial diversity. Furthermore, it will discuss the challenges associated with beampattern design and present state-of-the-art techniques and algorithms employed to optimize these patterns.

By comprehensively examining the intricacies of MIMO communication beampattern design, this research aims to contribute to the advancement of wireless communication systems. The outcomes of this study can potentially pave the way for future developments in MIMO technology, enabling the realization of efficient and reliable wireless networks that cater to the ever-growing demands of modern communication applications.

The study is divided into six sections as follows:

Section I provides a brief overview of MIMO communication.

Section II examines the use of covariance-based beamforming in MIMO communication.

Section III discusses the utilization of a model that concentrates the transmitter power at known desired locations.

Sections IV and V present an analysis of an algorithm for beam pattern design for both linear and planar arrays. These designs are compared with an ideal beamforming approach, and numerical results are provided.

Section VI focuses on the conclusion, and the paper includes references at the end.

## II. COVARIANCE BASED METHOD FOR MIMO COMMUNICATION BEAMPATTERN DESIGN

Let's consider a collection of N transmitter antennas placed at known positions in a spherical coordinate system along the z-axis. These antennas are aligned along the z-axis and are driven by specific signals on a carrier frequency $f_c$ or with a wavelength of $\lambda$. Each antenna generates a signal in the far field at a particular point in space, characterized by distance $d$ and direction $Ar(\theta, \phi)$ from the antenna.

The complex envelope of the total radiated signal at this point and in discrete form is given by Equation (1), where $EC_i(n, d, \theta, \phi)$ represents the signal generated by the i-th antenna.

$$EC_i(n, d, \theta, \phi) = \frac{1}{Kd} y_i\left(n - \frac{d}{c}\right) e^{j\left(\frac{2\pi}{\lambda}\right) P_i^T Ar(\theta,\phi)} \quad (1)$$

Where in this equation, $K$ is sphere surface constant equal to $\sqrt{4\pi}$, $y_i$ is complex envelope of each antenna transmitted signal and $P_i$ is position of i-th antenna.

In the far field, these signals and their powers combine linearly. The resulting combined signal from all the transmitted signals in the far field can be expressed as Equation (2).

$$EC(n, d, \theta, \phi) = \sum_{i=1}^{N} EC_i(n, d, \theta, \phi)$$

$$= \frac{1}{Kd} \sum_{i=1}^{N} y_i\left(n - \frac{d}{c}\right) e^{j\left(\frac{2\pi n z_i}{\lambda}\right) \sin(\theta)} \quad (2)$$

The resulted power of the combined signals which will be delivered to the users through the communication channel is given by Equation (3), which is a sum of cross-correlations between the transmitted signals [2].

$$P(\theta, \phi) = \frac{1}{K^2} \sum_{k=1}^{N} \sum_{l=1}^{N} R_{kl} e^{\frac{j2\pi}{\lambda}(z_k - z_l)\sin(\theta)} \quad (3)$$

The cross-correlation between two signals is defined as $R_{kl}$ in Equation (4).

$$R_{kl} = <y_k(t) y_l^*(t)> \quad (4)$$

By defining an steering vector of $s(\theta)$ in Equation (5) represents the conjugate transpose of the antenna response vector at θ, the power density $P(\theta, \phi)$ which is written in (3), can be expressed as Equation (6).

$$s(\theta) = \left[ e^{j\left(\frac{2\pi z_1}{\lambda}\right)\sin(\theta)}, \ldots, e^{j\left(\frac{2\pi z_N}{\lambda}\right)\sin(\theta)} \right]^T \quad (5)$$

$$P(\theta, \phi) = \frac{1}{4\pi} s^*(\theta) R s(\theta) \quad (6)$$

Equation (6) represents the power density P(θ, ɸ) for the users in terms of the cross-correlation matrix R represented in (7).

$$R = \sum_{k=1}^{N} \sum_{l=1}^{N} s_k(t) s_l^*(t) \quad (7)$$

These equations describe the desired beampattern generated by the cross-correlation matrix. In the following examples, we illustrate different beampatterns produced by such a matrix. Figure 1 shows the beampatterns of a 10-element uniform linear array (ULA) with half-wavelength spacing, generated by different signal cross-correlation matrices (8), (9), and (10).

It is important to note that these figures represent the directional characteristics of the array and provide information about the power distribution in different directions.

$$\begin{bmatrix} 1 & \cdots & 1 \\ \vdots & \ddots & \vdots \\ 1 & \cdots & 1 \end{bmatrix} \quad (8)$$

$$\begin{bmatrix} 0.8^0 & \cdots & 0.8^9 \\ \vdots & \ddots & \vdots \\ 0.8^9 & \cdots & 0.8^0 \end{bmatrix} \quad (9)$$

$$\begin{bmatrix} 1 & \cdots & 0 \\ \vdots & \ddots & \vdots \\ 0 & \cdots & 1 \end{bmatrix} \quad (10)$$

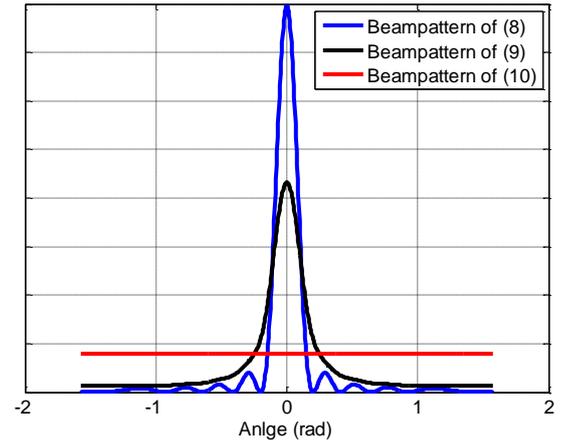

**Fig**. 1. Beampattern respect to (7). The blue one is corresponds to cross-correlation matrix of (8), The black one is corresponds to cross-correlation matrix of (9) and the red one is corresponds to cross-correlation matrix of (10)

In MIMO communication, the signal cross-correlation matrix usually consists of complex values, except for the real-valued diagonal elements. However, in the case of conventional communication, all transmitter signals are correlated with each

## III. MAXIMUM POWER DESIGN FOR KNOWN USER LOCATION

In this new task, we will be examining a set of antennas in a planar array. Our objective is to assess the effectiveness of the method described in reference [3] for maximizing the total power of probing signals at user's locations.

To summarize, we have K users of interest, with their positions denoted as $\{U_k\}_{k=1}^{K}$. The combined power of the all the antenna's signals at these user locations is determined by equation (11).

$$\sum_{k=1}^{K} s^*(\theta_k) R s(\theta_k) = tr\left(R \sum_{k=1}^{K} s(\theta_k) s^*(\theta_k)\right) \triangleq tr(RZ) \quad (11)$$

Where

$$Z = \sum_{k=1}^{K} s(\theta_k) s^*(\theta_k) \quad (12)$$

Now as in reference [??] one can aims to maximize equation (11) at the desired locations, while adhering to certain constraints. So the optimization problem can be formulated as maximizing $tr(RZ)$ subject to maximum transmitted power, as depicted in equation (13).

$$\max_{Z} tr(RZ) \; subject\; to$$
$$tr(R) = P_t \; (Max\; TX\; Power)$$
$$R \geq 0$$
$$(13)$$

This equation (13) is a well know linear algebra optimization problem and has solution like the authors in reference [??] have shown for linear arrays and discovered that the optimal value of R is given by equation (14), where **v** is the eigenvector associated with the highest eigenvalue of **Z**.

$$R = vv^* \quad (14)$$

In this paper, our goal is to explore this problem for planar array transmitter sets and assess the accuracy of this method. The following section will present numerical examples to illustrate this.

## IV. LINEAR ARRAYS

As previously mentioned, the issue of linear arrays has been explored by authors in [3]. This section aims to examine their findings. For instance, let's consider a linear array with 50 transmitter elements positioned along the z-axis. Figure 2 illustrates a random arrangement of the users of interest, with the objective of concentrating the transmitter power around them along the y and z axes. If we analyze this problem using the approach outlined in section III, the resulting beampatterns will resemble those shown in Figure 3.

It is worth noting that in Figure 3, the ideal beampattern corresponds to the beampattern where the transmitted covariance matrix is equivalent to Z in equation (12). Additionally, since the linear array is aligned with the z-axis, it can only differentiate between users located at different positions along this axis.

From Figure 3, it is evident that the proposed method failed to evenly focus its transmitted power around all users. This represents a drawback of the method.

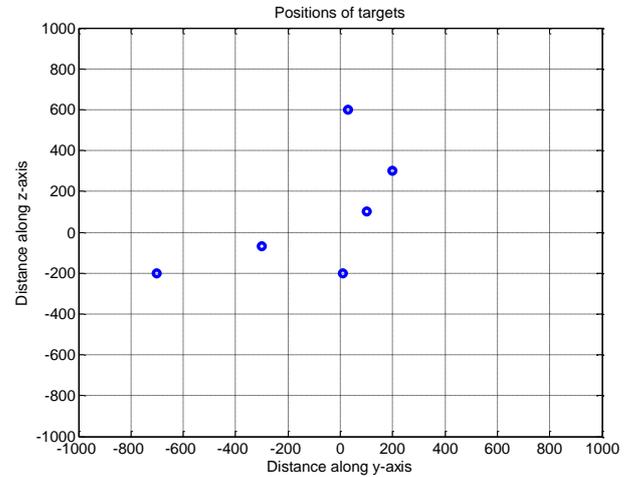

**Fig**. 2. Y and Z values of users placement. X value is equal to 100 km

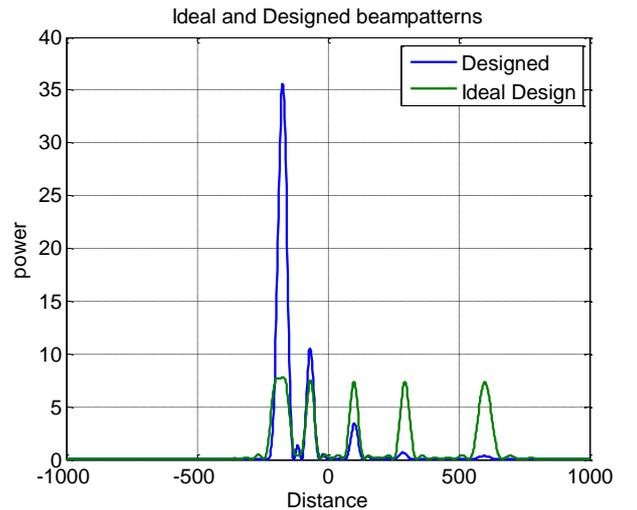

**Fig**. 3. comparision of designed and ideal beampatterns in linear arrays

## V. PLANAR ARRAYS

In this section, it is important to explore the precision of the proposed beampattern design method for a plannar array. To assess the accuracy, we will examine various examples and figures presented in this paper. It should be noted that the transmitter antennas are assumed to be positioned along the y and z axis throughout.

### A. Square constellation

In this subsection, we assume the presence of a flat array with a square pattern and a total of 400 transmitter elements (20×20 array) as shown in Figure 4.

From the illustration, it is evident that the transmitter elements of this array are evenly distributed along the y and z axes, with a spacing of half a wavelength between them.

In this scenario, we consider a situation where there are six users of interest located 100 km away from the origin along the x-axis, as depicted in Figure 5.

It is important to note that, going forward, the user placement shown in Figure 5 will be used for simulations involving all planar arrays constellations.

Since all elements of vector **s** in (5) have an amplitude of 1, equation (7) reveals that the matrix R, which the cost function, will take the following form:

$$R_{max} = \sum_{k=1}^{\tilde{k}} s(\theta_k)s(\theta_k) \qquad (15)$$

In this equation, as said before, k represents the number of users. While the above expression maximizes power around the uses, it may not fulfill the requirement of being a cross-correlation matrix of signals, as stated in equation (13). The cross-correlation matrix of signals must be non-negative definite. We will utilize this matrix for result comparison, and the resulting beampattern from this cross-correlation matrix of transmitted signals will be referred to as the ideal beampattern.

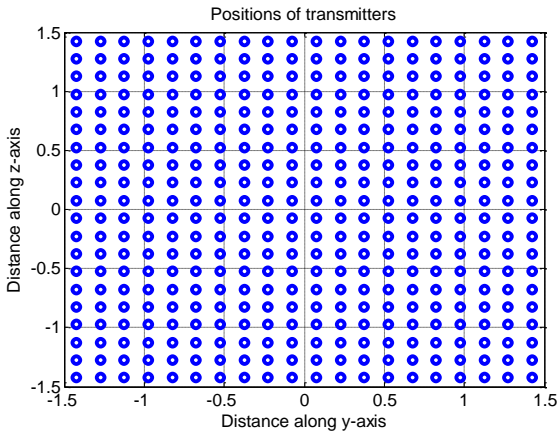

Fig. 4. Placement of transmitter antenna array

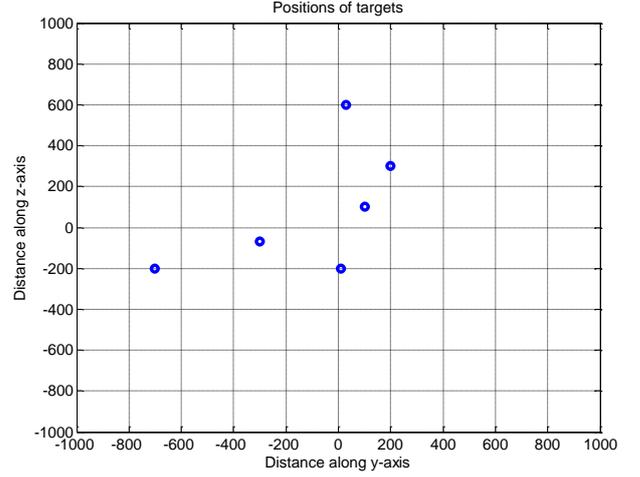

Fig. 5. Y and Z values of users placenent. X value id equal to 100 km

Figure 6 and Figure 7 display the 3D and top view of the ideal beampattern in this case, respectively. These figures demonstrate that the ideal beampattern successfully resolves all six users and allocates equal power to each of them.

Now, Figure 8 and Figure 9 present the designed beampatterns based on equation (14). Notably, Figure 8 illustrates the 3D designed beampattern, while Figure 9 shows the same beampattern from a top view.

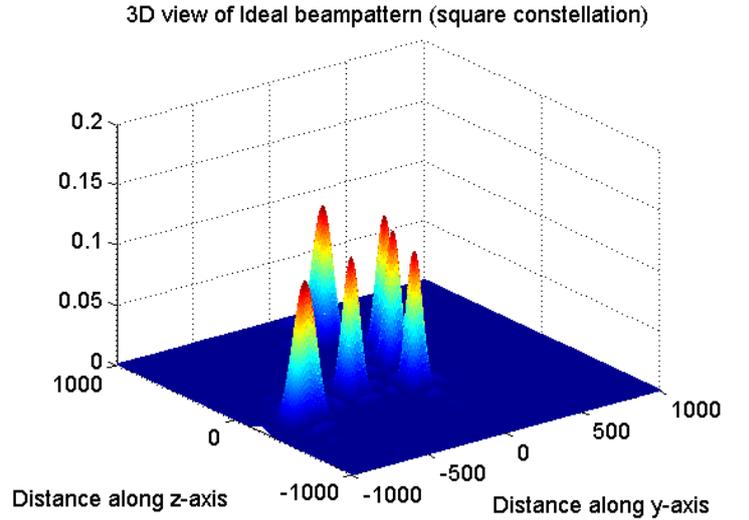

Fig. 6. 3D view of ideal beampattern of square constellation

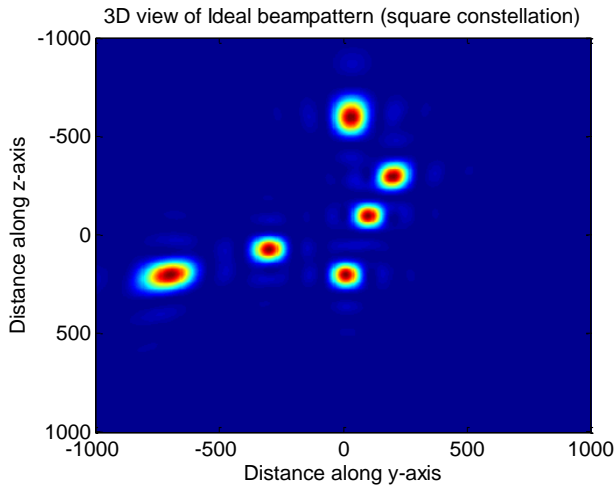

Fig. 7. top view of ideal beampattern of square constellation

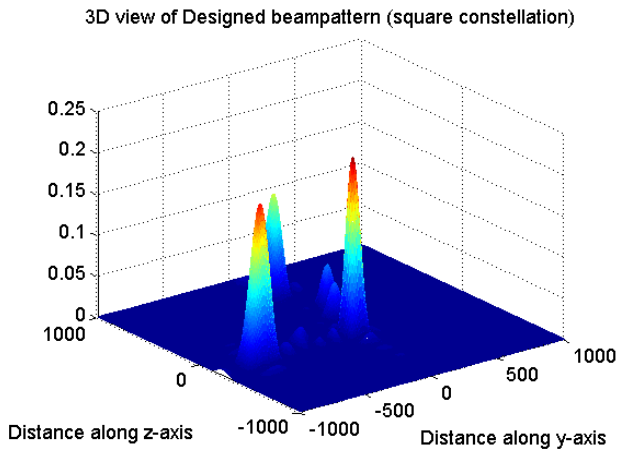

Fig. 8. 3D view of designed beampattern of square constellation related to (14)

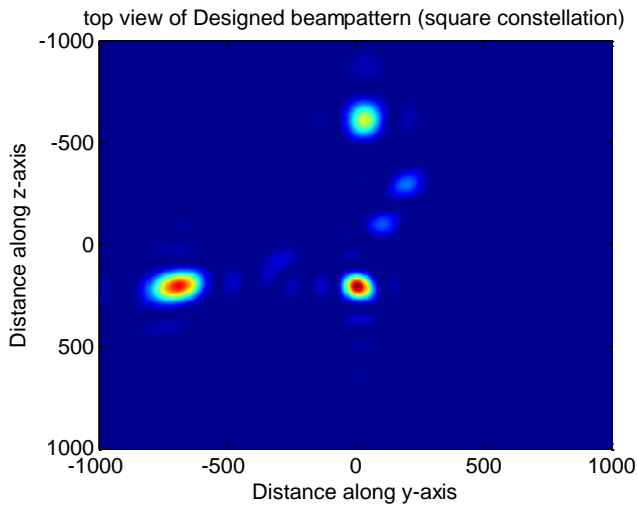

Fig. 9. 3D view of designed beampattern of square constellation related to (14)

As observed from this figure, this algorithm successfully identifies three out of the total six users with concentrated power around them. However, three users are missed, and the power allocated to these users is not equal. In other words, this algorithm does not provide the designer with the ability to control the power to focus on one or more desired directions, which may be more important.

### B. Circle constellation

In this subsection, we assume the presence of a planar array with a circular arrangement consisting of a total of 400 transmitter elements, as shown in Figure 10.

From the illustration, it can be observed that the transmitter elements in this array are evenly distributed along the y and z axes, with a spacing of half a wavelength between them.

Figures 11 and 12 display the three-dimensional and top view representations of the ideal beampattern for this case, respectively. These figures reveal that the ideal beampattern successfully resolves all six users and allocates equal power to each of them.

It is worth noting that the ideal beampattern of the circular arrangement, as shown in these figures, bears resemblance to the ideal beampattern of a square arrangement. However, they differ in terms of sidelobe level and half-power beadwidth. Consequently, the ideal beampatterns for each arrangement will be depicted separately.

Moving on to Figures 13 and 14, they illustrate the designed beampatterns obtained from equation (14). Figure 13 showcases the three-dimensional designed beampattern, while Figure 14 provides a top view of the same beampattern.

As evident from these figures, this algorithm successfully identifies four out of the six users with a strong concentration of power around them. However, approximately two users are missed, albeit at a weaker peak level in the transmitted beampattern. Furthermore, the powers assigned to these users are not equal to each other. In other words, this algorithm does not offer the designer control over power allocation to focus it in specific desired directions, which may be more crucial.

An important observation to make here is the disparity between the designed beampatterns of the circular and square constellations. As illustrated in these figures, the proposed algorithm performs better in a circular constellation compared to a square one.

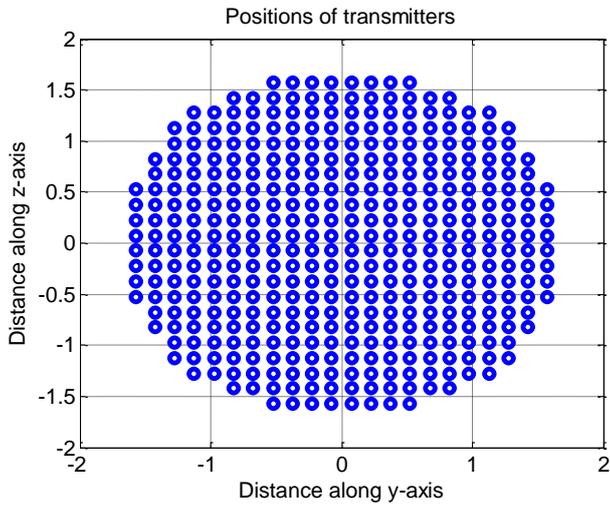

**Fig**. 10.  Placement of transmitter antenna array

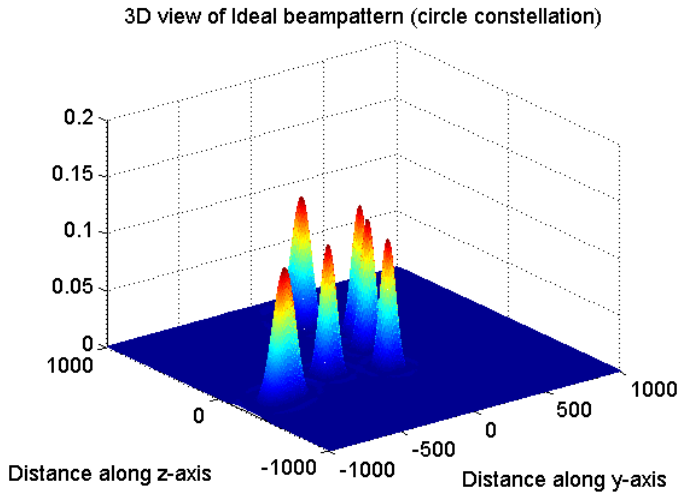

**Fig**. 11.  3D view of ideal beampattern of circle constellation

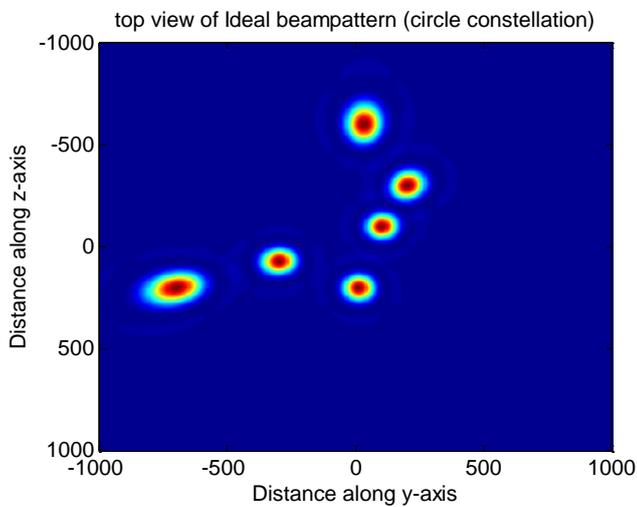

**Fig**. 12.  top view of ideal beampattern of circle constellation

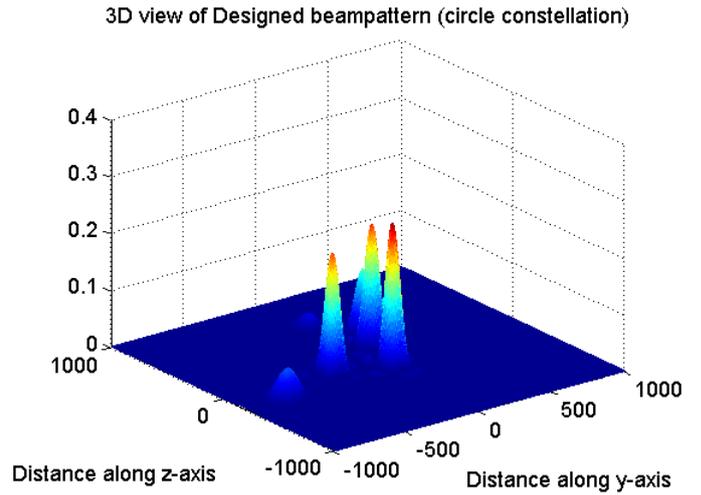

**Fig**. 13.  3D view of designed beampattern of circle constellation related to (14)

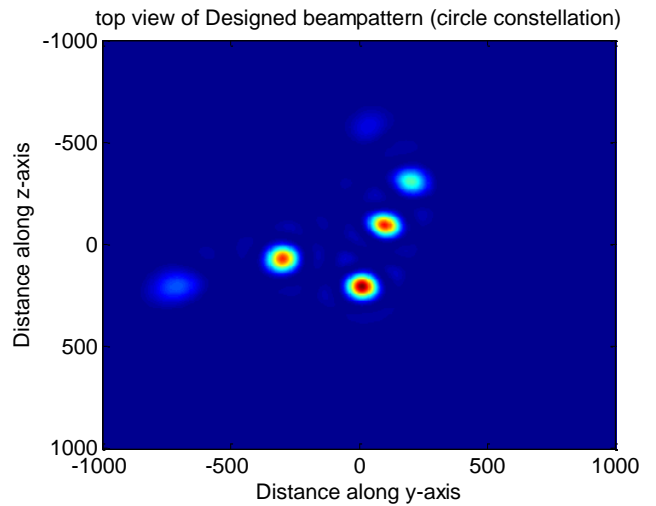

**Fig**. 14.  top view of designed beampattern of circle  constellation related to (14)

*C.  Hexagonal constellation*

In this subsection, we will consider a planar array with a hexagonal arrangement consisting of 400 transmitter elements, as shown in Figure 15. It should be noted that this hexagonal constellation is an approximate representation of a circular constellation, which is sometimes used in practice as an alternative.

Figure 15 reveals that the transmitter elements of this array are evenly distributed along the y and z axes, with a spacing of half a wavelength between them.

Figures 16 and 17 illustrate the 3D and top view of the ideal beampattern for this case, respectively. These figures demonstrate that the ideal beampattern successfully separates all six users and allocates equal power to each of them.

It should be noted that the ideal beampattern of the hexagonal constellation, as shown in these figures, resembles the ideal beampattern of previous constellations. However, there are differences in terms of sidelobe level and half power beadwidth. Therefore, the ideal beampatterns of each constellation will be depicted separately.

Now, let's examine Figures 18 and 19, which display the designed beampatterns obtained from equation (14). Figure 18 presents the 3D designed beampattern, while Figure 19 provides a top view of the same beampattern.

As observed in Figure 19, this algorithm successfully identifies four out of the six users with concentrated power around them. However, two users were missed, and the power allocated to each user is not equal. In other words, this algorithm lacks the ability to control the power distribution and focus it on specific desired directions, which may be more important.

Based on these simulations, it is evident that the hexagonal constellation performs better than the square constellation but worse than the circular constellation. However, when considering the results of both the square and hexagonal constellations, it becomes apparent that these constellations complement each other in maximizing power around the users' locations.

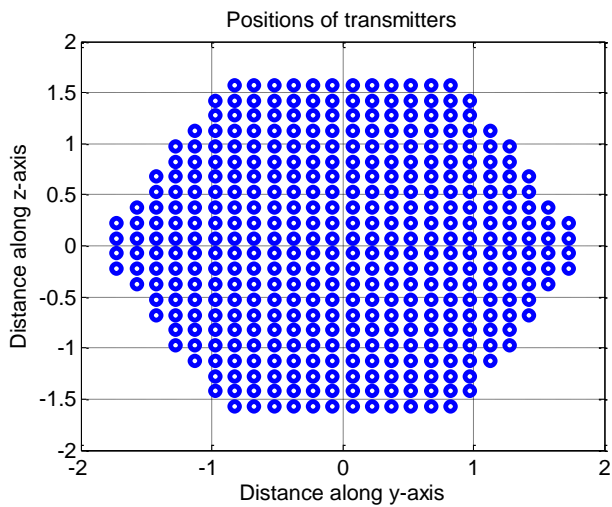

**Fig**. 15.  Placement of transmitter antenna array

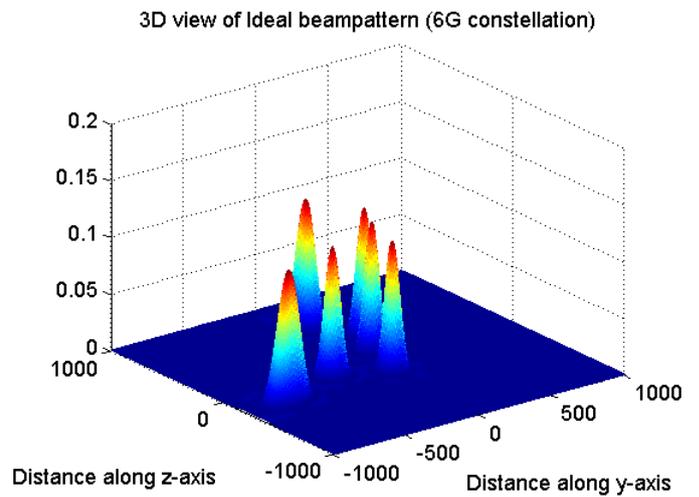

**Fig**. 16.  3D view of ideal beampattern of hexagonal constellation

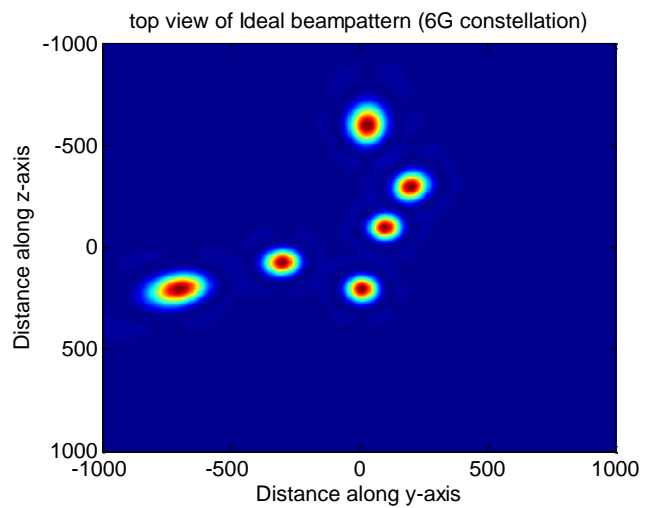

**Fig**. 17.  top view of ideal beampattern of hexagonal constellation

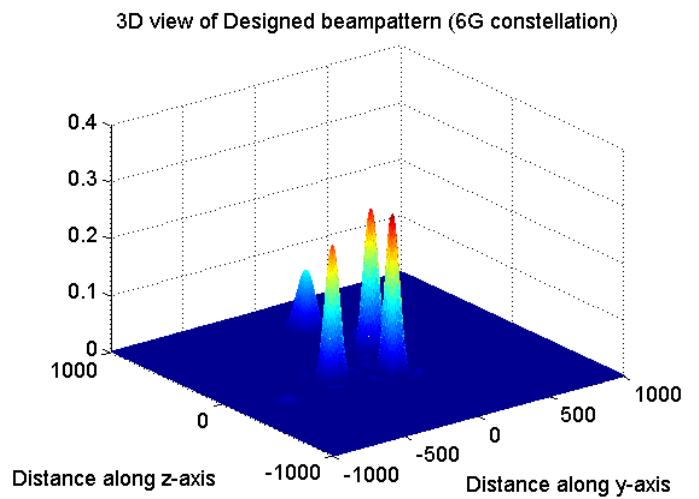

**Fig**. 18.  3D view of designed beampattern of hexagonal  constellation related to (14)

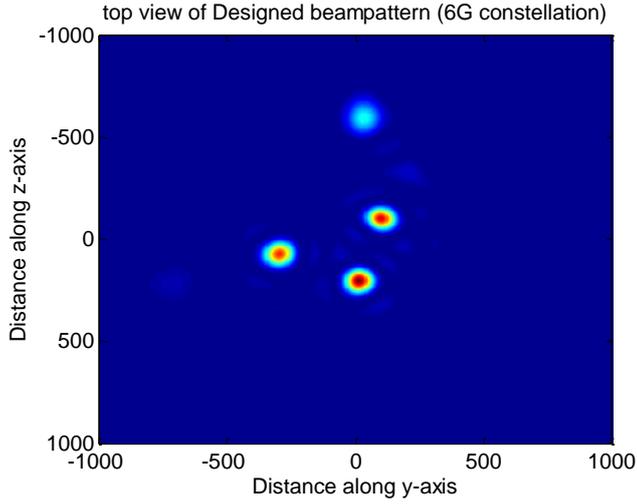

**Fig**. 19. top view of designed beampattern of hexagonal constellation related to (14)

### D. Spiral constellation

In this subsection, we assume the presence of a planar array with a spiral pattern and a total of 400 transmitter elements, as shown in Figure 20.

In certain applications and for specific purposes, one may opt to use a spiral pattern as it is very well-known in the concept of antennas. The equation for the spiral is given by:

$$r = a.e^{b\theta} \quad (16)$$

Here, $r$ represents the radius from the center and θ, in radians, represents the angle from the y-axis in Figure 20.

It is important to note that, for the simulations in this subsection, we make the following assumptions:

$$a = 0.15 \quad (17)$$

$$b = 0.1 \quad (18)$$

As depicted in this figure, the transmitter elements of this array are not uniformly distributed along the y and z axes. In other words, their exact locations along the y and z axes can be determined using the following equations:

$$y = r.\cos\theta \quad (19)$$

$$z = r.\sin\theta \quad (20)$$

Figure 21 and Figure 22 display the 3D and top views of the ideal beampattern for this case, respectively. These figures demonstrate that the ideal beampattern successfully resolves all six users and allocates equal power to each of them.

It is worth noting that the ideal beampattern of the spiral pattern, as shown in these figures, is similar to the ideal beampatterns of previous constellations. However, they differ in terms of sidelobe level and half power beamwidth. For instance, in this case, the sidelobe levels are higher compared to previous cases. Therefore, we will depict the ideal beampatterns of each constellation separately.

Now, Figure 23 and Figure 24 illustrate the designed beampatterns obtained from equation (14). Figure 23 represents the 3D designed beampattern, while Figure 24 provides a top view of the same beampattern.

As observed in these figures, this algorithm successfully identifies four out of the six users, concentrating power effectively around them. However, two users have been missed, and the power allocated to these users is not equal. In other words, this algorithm does not provide the designer with the ability to control and focus the power towards specific directions that may be more important.

In comparison to previous results, although this constellation also misses some users like the latter results, it exhibits higher sidelobe levels around the missed users, which can aid in detecting and maximizing the transmitted power towards those users.

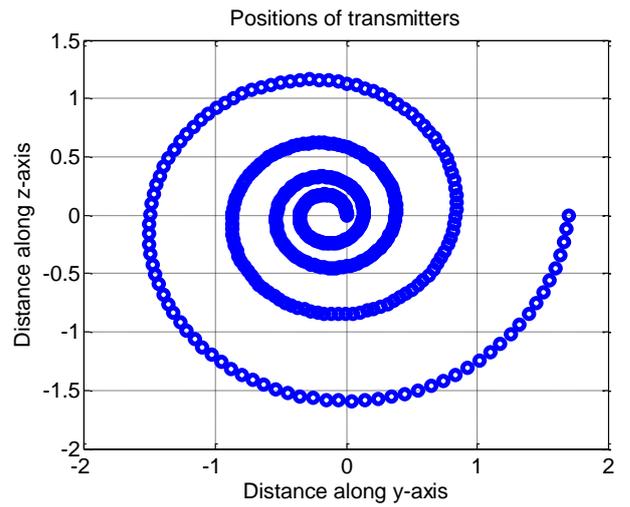

**Fig**. 20. Placement of transmitter antenna array

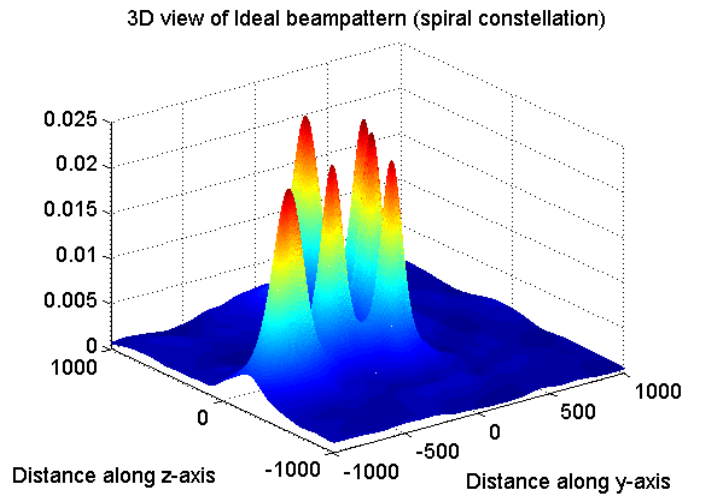

**Fig**. 21. 3D view of ideal beampattern of spiral constellation

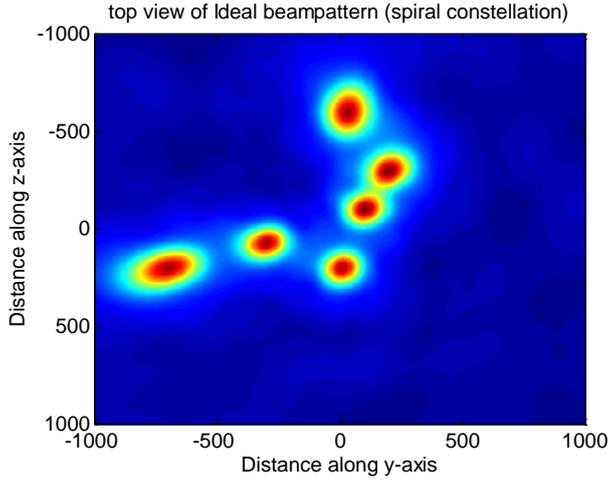

**Fig**. 22. top view of ideal beampattern of spiral constellation

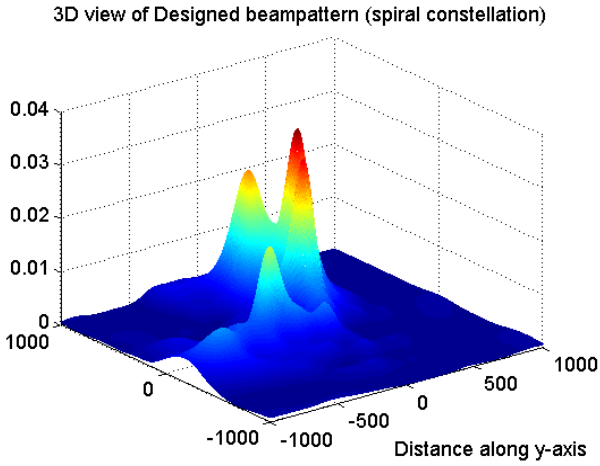

**Fig**. 23. 3D view of designed beampattern of spiral constellation related to (14)

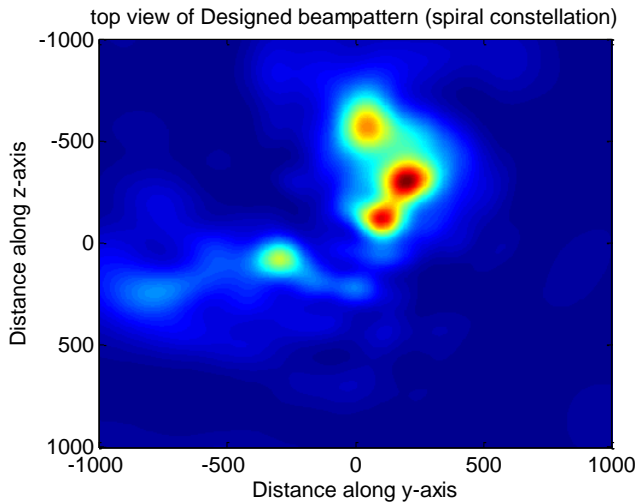

**Fig**. 24. top view of designed beampattern of spiral constellation related to (14)

*E. Archimedes spiral constellation*

In this subsection, we assume the presence of a planar array with an Archimedes spiral arrangement consisting of a total of 400 transmitter elements, as depicted in Figure 25 and Figure 26.

In certain specific applications and for particular purposes, it is possible to utilize a spiral arrangement. The equation for the spiral is as follows:

$$r = a.\theta^{\frac{1}{n}} \quad (21)$$

Where r represents the radius from the center and θ, in radians, represents the angle from the y-axis in Figure 25 and Figure 26.

Here, we will consider this arrangement for two different values of n, namely n=1 and n=3. Therefore, it is important to note that in all simulations conducted in this section, the first figure corresponds to n=1 while the second figure corresponds to n=3. Additionally, for the sake of simplicity, the following assumptions have been made:

$$a = 0.08 \quad for \quad n = 1 \quad (22)$$
$$a = 0.28 \quad for \quad n = 3 \quad (23)$$

As observed from these figures, the transmitter elements of this array are not uniformly distributed along the y and z axes. In other words, their precise locations along the y and z axes can be determined using the following equations:

$$a = 0.08 \quad for \quad n = 1 \quad (22)$$
$$a = 0.28 \quad for \quad n = 3 \quad (23)$$

Figure 27, Figure 28, Figure 29, and Figure 30 display the three-dimensional and top views of the ideal beampattern for this case, respectively. As evident from these figures, the ideal beampattern successfully separates all six users and allocates equal power to each of them.

It should be noted that the ideal beampattern of the Archimedes spiral arrangement, as seen in these figures, is similar to that of previous constellations. However, they are not identical and differ in terms of sidelobe level and half power beamwidth. For instance, in this case, the sidelobe levels are higher compared to uniform arrangements. Therefore, the ideal beampatterns of each constellation will be depicted separately.

Now, Figure 31, Figure 32, Figure 33, and Figure 34 demonstrate the designed beampatterns derived from equation (14). It is worth mentioning that Figure 31 and Figure 32 present the three-dimensional designed beampattern, while Figure 33 and Figure 34 provide a top view of the same beampattern.

As evident from these figures, the results obtained from such an arrangement may even surpass those achieved with uniform arrays when employing the proposed algorithm in equation (14) and these constellations are able to focus their power toward 5 users.

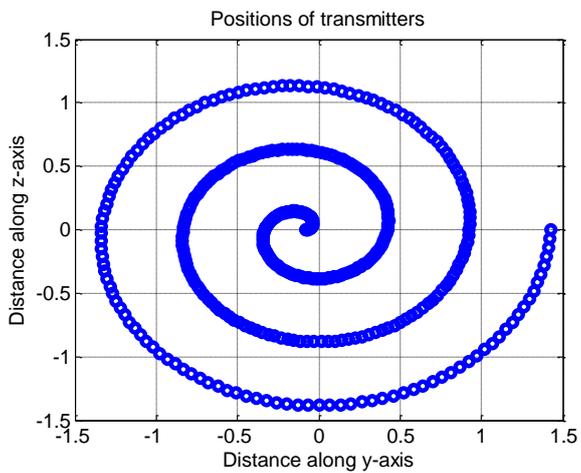

**Fig**. 25. Placement of transmitter antenna array (Archimedes spiral, n=1)

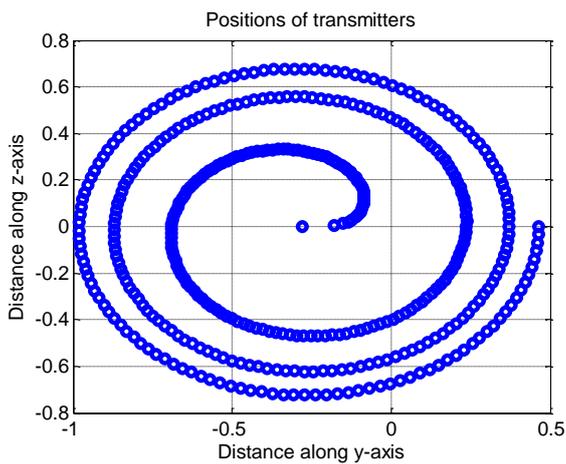

**Fig**. 26. Placement of transmitter antenna array (Archimedes spiral, n=3)

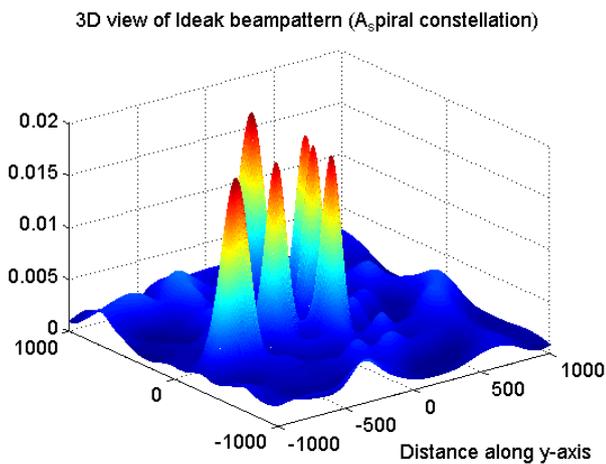

**Fig**. 27. 3D view of ideal beampattern of Archimedes spiral constellation (n=1)

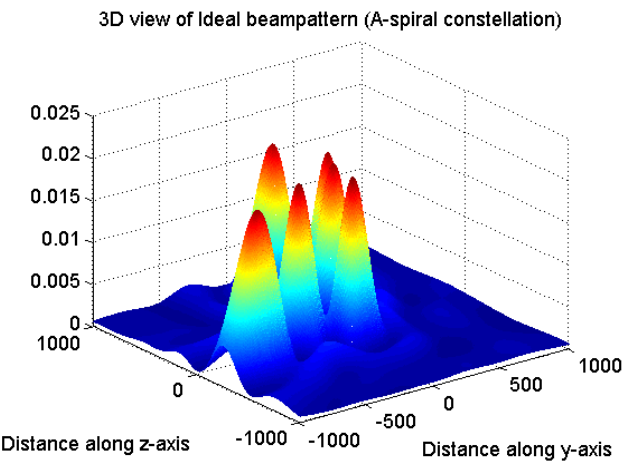

**Fig**. 28. 3D view of ideal beampattern of Archimedes spiral constellation (n=3)

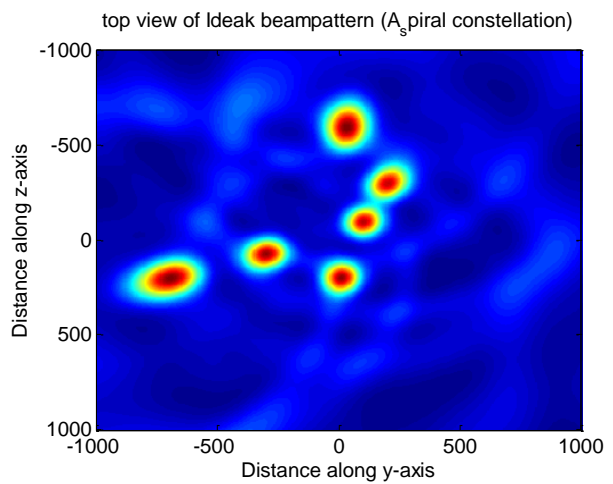

**Fig**. 29. top view of ideal beampattern of Archimedes spiral constellation (n=1)

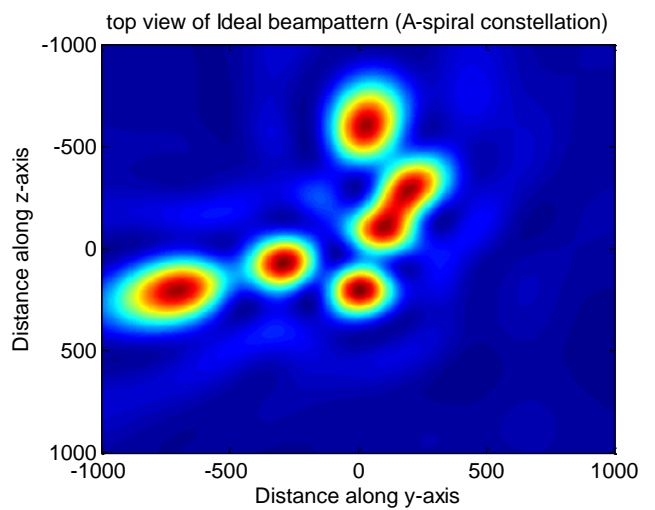

**Fig**. 30. top view of ideal beampattern of Archimedes spiral constellation (n=3)

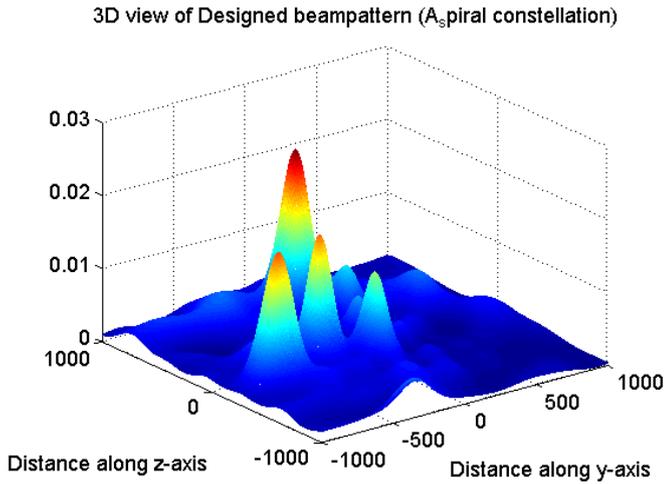

**Fig**. 31. 3D view of designed beampattern of Archimedes spiral constellation (n=1) related to (14)

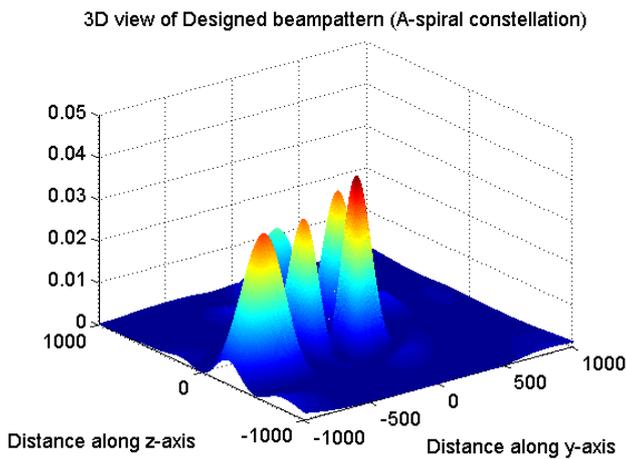

**Fig**. 32. 3D view of designed beampattern of Archimedes spiral constellation (n=3) related to (14)

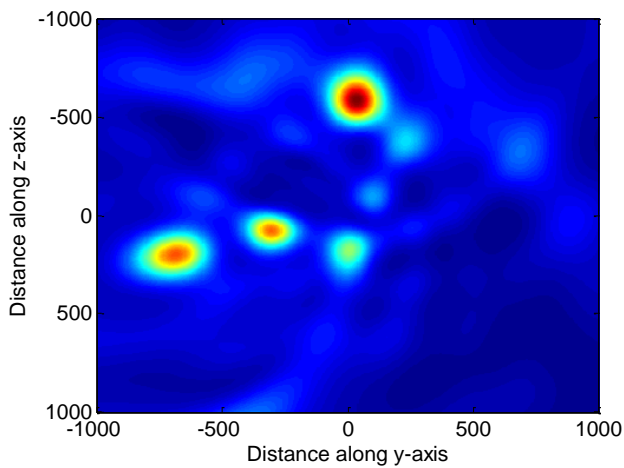

**Fig**. 33. top view of designed beampattern of Archimedes spiral constellation (n=1) related to (14)

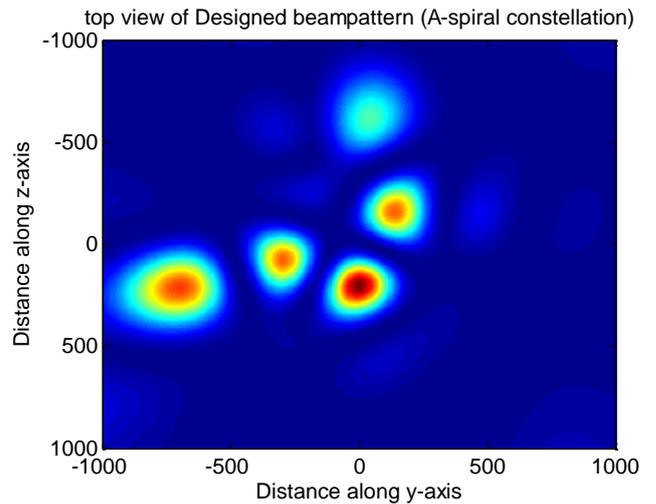

**Fig**. 34. top view of designed beampattern of Archimedes spiral constellation (n=3) related to (14)

## VI. CONCLUSION

In conclusion, the analysis of 3D MIMO communication beamforming in linear and planar arrays has demonstrated its immense potential in improving wireless communication performance. As this technology continues to evolve and mature, it holds great promise for enhancing network capacity, coverage, and overall user experience in diverse environments.

It is showed that an approach which had good results for a linear array, here in planar array didn't have good results necessarily and it should be opted based on the geographical of a city or the place where the antennas are located. As futures work one can consider other beam-forming approaches in planar arrays and also one can look for approaches which provide suitable result for planar arrays as well as linear arrays.

.